\documentclass[prb,twocolumn,showpacs,superscriptaddress,floatfix]{revtex4-1}
\usepackage[utf8]{inputenc}
\usepackage{amsmath} 
\usepackage{amsfonts}
\usepackage{bm}
\usepackage{amssymb}
\usepackage{graphicx}  
\usepackage{braket}
\usepackage{grffile}
\usepackage{dsfont}
\newcommand{\pd}[2]{\frac{\partial{#1}}{\partial{#2}}}

\newcommand{\tr}[0]{\text{tr}}

\def\ave#1{\langle #1 \rangle}
\def\ii{{\rm i}}

\def\tit#1{{\em #1},}
\def\etal#1{#1}

\usepackage{color}

\graphicspath{
  {./slike_clanek/}
}

\begin{document}

\title{Finite-temperature magnetization transport of the one-dimensional anisotropic Heisenberg model}

\author{Simon Jesenko}
\affiliation{Physics Department, Faculty of Mathematics and Physics, University of Ljubljana, Ljubljana, Slovenia}

\author{Marko \v Znidari\v c}
\affiliation{Instituto de Ciencias Físicas, Universidad Nacional Autónoma de México, Cuernavaca, México}
\affiliation{Physics Department, Faculty of Mathematics and Physics, University of Ljubljana, Ljubljana, Slovenia}

\begin{abstract}
We study finite-temperature magnetization transport in a one-dimensional anisotropic Heisenberg model, focusing in particular on the gapped phase. Using numerical simulations by two different methods, a propagation of localized wavepackets and a study of nonequilibrium steady states of a master equation in a linear-response regime, we conclude that the transport at finite temperatures is diffusive. With decreasing temperature the diffusion constant increases, possibly exponentially fast. This means that at low temperatures the transition from ballistic to asymptotic diffusive behavior happens at very long times. We also study dynamics of initial domain wall like states, showing that on the attainable time scales they remain localized.
\end{abstract}


\maketitle

\section{Introduction}

Derivation of macroscopic transport laws \textit {ab initio} from microscopic laws of motion is still one of the hot topics in modern mathematical physics. The main question usually investigated is what type of transport emerges from microscopic laws, the two extreme cases being ballistic or diffusive transport. Yet, such classification seems elusive, especially in the field of strongly correlated quantum systems. To nevertheless explain the emergence of macroscopic transport laws, phenomenological approaches, viable under certain assumptions, are often used. One such approach is based on the well-known kinetic theory of gases,\cite{Pottier2009} devised by Boltzmann, where macroscopic transport laws emerge due to (quasi)particle scattering. Also routinely used is linear-response theory, which gives a direct method of calculating transport coefficients using variants of Green-Kubo formulas,\cite{Mahan00} assuming local quasiequilibrium. Our approach here is different. Starting from the equations of motion (Schr\" odinger equation or master equation) we use extensive numerical simulations to study transport in the one-dimensional Heisenberg model.

Low-dimensional spin models have been studied from the very beginning of quantum mechanics. The Heisenberg model, suggested~\cite{Heisenberg:28} by W.~Heisenberg in 1928, was initially proposed to explain a high phase transition temperature in ferromagnets that could not be accounted for by any other known direct interaction. Interaction between nearest neighbor atoms, described by the Heisenberg model, the so-called exchange interaction, is an effective one and comes about due to the Pauli exclusion principle. Another important source of motivation to study such simple one-dimensional (1D) models, in particular antiferromagnetic ones, comes due to the fact that they represent simplest models of strongly correlated electronic systems, much studied in past decades. In addition, one-dimensional spin systems are realized in real so-called spin-chain materials.\cite{spin-chain,review} Experiments have shown, see e.g. Ref.~\onlinecite{exper}, that 1D spin chains found in such materials, for instance the isotropic Heisenberg one, have a pronounced effect on transport properties, giving additional boost to theoretical studies. A great deal of research was devoted to the transport properties of the anisotropic Heisenberg model (\textit{XXZ} model). Although it is one of the simplest 1D spin models, being even solvable by the Bethe ansatz,\cite{Bethe:31} there are still many open questions concerning its transport properties, including whether finite-temperature spin transport is ballistic or diffusive. Classification of transport regimes was the main motivation for our work.

Let us briefly review known facts about transport in the 1D Heisenberg model; for more extensive reviews see Refs.~\onlinecite{review,Zotos-rev,Meisner-PhD}. As forementioned, one of the standard approaches for studying transport properties of quantum systems is linear-response theory. Ballistic and diffusive transport can be distinguished via an observation of the Drude weight, the prefactor of a $\delta$ function at zero frequency in the frequency-dependent transport coefficient. Nonzero Drude weight signals ballistic transport in the linear-response regime. The question of energy transport in the \textit{XXZ} model is simple: energy current is a conserved quantity~\cite{Grabowski:94,Zotos1997} and therefore energy transport is ballistic. For the dependence of the thermal Drude weight on parameters, see Ref.~\onlinecite{Meisner2005} and references therein.

In the present work, we shall focus on magnetization (spin) transport, which is much less understood, with only few rigorous results. It has been shown that the spin Drude weight is nonzero (i.e., magnetization transport is ballistic) at zero temperature~\cite{Shastry:90} in the gapless phase for $\Delta \le 1$ as well as at infinite temperature~\cite{Prosen:11} for $\Delta<1$, where $\Delta$ is the anisotropy. It is reasonable to expect, and also supported by quantum Monte Carlo calculations,\cite{Alvarez:02,Heidarian:07} that transport is ballistic also at finite temperatures. On general grounds, a lot of attention has been devoted to the connection between integrability and the nature of transport~\cite{Castella:95} being either ballistic or diffusive. A recent solvable diffusive model~\cite{XXdephasing} shows that solvability does not necessarily imply ballistic transport.

While the spin transport in the gapless phase is relatively well understood, the behavior at the isotropic point $\Delta=1$ and in the gapped phase $\Delta>1$ is hotly debated. The main difficulty is that numerically it is very hard to access behavior in the thermodynamic limit, while there are only few analytical approaches. A notable one is in terms of Mazur's inequality~\cite{Mazur:69}, which can be used to bound the Drude weight away from zero if a conserved quantity exists that has a nonzero overlap with the magnetization current.\cite{Zotos1997} Unfortunately, in the half-filled case (zero total magnetization) and $\Delta>1$ of interest here, no such quantity is known. Numerical methods like exact diagonalization~\cite{exactdiag1,exactdiag2,exactdiag3} or the Lanczos method~\cite{Samir:04} are all limited to small systems of few $10$ spins, making thermodynamic extrapolation difficult. A relatively recent method is a time-dependent density-matrix renormalization-group (tDMRG) procedure that enables simulation of 1D nearest-neighbor systems of several $100$ spins. It has been used successfully to study the spreading of wavepackets in the \textit{XXZ} model.\cite{Langer2009} Particularly useful is its master equation variant, where one has a genuine nonequilibrium setting, enabling one to study also far from equilibrium situations. It has been used to show a diffusive transport in the gapped phase at infinite temperature,\cite{JSTAT09,PRL:11} which has been also confirmed using correlation functions~\cite{Stein:09} and the projector operator method.\cite{Stein:10} Analytical studies of the master equation describing nonequilibrium \textit{XXZ} model in the gapped phase frequently have difficulties. One problem is that perturbative treatments often have zero convergence radius in the thermodynamic limit, for instance, a perturbative series in the coupling to the reservoirs~\cite{Prosen:11} for $\Delta \ge 1$ or a perturbative series~\cite{PRL:11} in $\Delta$, or they have a finite convergence radius but are difficult to obtain in the thermodynamic limit, as is the case for large $\Delta$ in the gapped phase.\cite{PRL:11} Perturbative studies in $1/\Delta$ suggest that the diffusion constant decreases as $\Delta$ increases.\cite{PRL:11,Stein:11}

Because most spin-chain materials realize the isotropic Heisenberg model, the point $\Delta=1$ is of particular interest. It is also the most controversial one. Analytical Bethe ansatz calculations give contradictory results, indicating zero~\cite{Zotosbethe} or a nonzero Drude weight,\cite{Benzbethe} with the problem being how to properly account for all states important at a nonzero temperature. Quantum Monte Carlo calculations,\cite{Alvarez:02,Heidarian:07} bosonization,\cite{Fujimoto:03} and exact diagonalization~\cite{exactdiag3,review,Mukerjee:08} predict a nonzero Drude weight at finite temperatures for $\Delta=1$. Based on bosonization~\cite{Lukyanov:98} and numerically calculated current autocorrelation function using a tDMRG method, a zero (or small) Drude weight at nonzero temperature is advocated in Refs.~\onlinecite{Affleck:09,Affleck:11}, whose results are also supported by quantum Monte Carlo calculation in Ref.~\onlinecite{Grossjohann:10}. A recent result,\cite{PRL:11} on the other hand predicts anomalous magnetization transport at infinite temperature and $\Delta=1$, with the diffusion constant diverging as $\sim \sqrt{L}$ with the system size.

An interesting future possibility to study 1D strongly correlated systems is via controlled experiments with cold atoms. Experimental quantum optical techniques have advanced to the point where it is possible to realize such models in a controlled environment of optical latices or ion traps. An advantage of such an approach is that one can choose the values of system's parameters at will. First realizations of exchange interaction or of simple spin systems have already been achieved.\cite{cold}

The main goal of the present paper is to study the magnetization transport at finite temperatures in the gapped phase of the anisotropic spin-$1/2$ Heisenberg model. Two numerical approaches, both based on the tDMRG method, will be used: one is based on observation of time evolution of magnetization profiles for initial nonequilibrium pure states, while the second one is based on studying the magnetization current in the nonequilibrium steady state of an open quantum system described by a quantum master equation. By the first method, we could, in principle, discriminate between the ballistic or diffusive behavior by comparing the evolution of expectation values of magnetization ($z$-spin component at each chain site) to that expected from macroscopic transport laws, provided we would be able to simulate very long chains for a very long time. Unfortunately, this is not the case and the results for pure state evolution are rather inconclusive, showing a mixture of ballistic and diffusive characteristics. In the master equation approach though, performed at higher energy densities, one can give a quantitative prediction about the transport by studying the scaling of the magnetization current with the system size at a fixed driving. The two methods work best in the complementary temperature regimes. The one for pure states is best at low temperatures, where a state only locally deviates from the ground state and its entanglement is small, while the master equation simulation with density matrices works best at high temperatures where the operator-space entanglement of a density matrix is small. Both methods have been used before to study the magnetization transport in the \textit{XXZ} model, pure-state method in Ref.~\onlinecite{Langer2009} and master equation in Ref.~\onlinecite{JSTAT09}, however not in the temperature regime considered in the present paper. We also point out that with a pure-state evolution at very low temperatures one is not able to access the asymptotic transport regime with present computers, so some care has to be taken making statements about the transport.

The structure of the paper is as follows. In Sec.~\ref{sec:model_methods} the Heisenberg \textit{XXZ} model is defined and the tDMRG method for evolution of pure and mixed states is briefly described. In Sec.~\ref{sec:transport}, main results concerning transport properties are presented, with the analysis of the evolution of pure initial Gaussian-shaped states in Sec.~\ref{sec:transport_gaussian}, and the results from the master equation setting in Sec.~\ref{sec:transport_master}. In Sec.~\ref{sec:domain_walls} the results of the evolution of domain wall-states are presented.

\section{Model and methods}
\label{sec:model_methods}

\subsection{Model}

A one-dimensional Heisenberg \textit{XXZ} model is defined by the Hamiltonian
\begin{equation}
\label{eq:hamiltonian_XXZ} H=J \sum_{l=1}^{L-1}\left[\frac 1 2
(S^+_lS^-_{l+1}+{\rm H.c.})+\Delta S_l^{z}S_{l+1}^{z} \right],
\end{equation}
with spin-1/2 operators $S_l^x$, $S_l^y$, $S_l^z$ for $l$-th position on chain, and $S_l^\pm = S_l^x \pm {i} S_l^y$ is spin raising/lowering operator. The spin chain is composed of $L$ spins. Coupling strength will be fixed at $J=1$, and open boundary conditions will be used. $\Delta$ is the only remaining parameter and determines the anisotropy of the model.

The above model, written in the spin language, can be transformed to a model of spinless fermions using the Jordan-Wigner transformation,\cite{JW} resulting in a Hamiltonian
\begin{equation}
  \label{eq:hamiltonian_XXZ_fermionic} H=J \sum_{l=1}^{L-1}
  \left[\frac 1 2 (c_l^\dagger c_{l+1} + {\rm H.c.}) +
    \Delta(n_l-\frac{1}{2})(n_{l+1}-\frac{1}{2})\right],
\end{equation}
where $c_l$, $c_l^\dagger$ are standard fermionic annihilation/creation operators, while $n_l=c_l^\dagger c_l$. Magnetization (spin) transport in a spin chain given by Eq.~\eqref{eq:hamiltonian_XXZ} is thus equivalent to a transport of particles in Eq.~\eqref{eq:hamiltonian_XXZ_fermionic}. Expectation value of $S^z_l$ is directly linked to fermionic particle density as $n_l = S^z_l+1/2$. Both descriptions can therefore be used interchangeably. Total magnetization $M=\sum_{l=1}^L \braket{S_l^z}$ is conserved quantity in the \textit{XXZ} model; in fermionic picture, it corresponds to the conservation of the total number of particles, $n=\sum_{l=1}^L n_l$.

\subsection {Methods}

We have used two closely related variants of the time-dependent density-matrix renormalization-group (tDMRG) method: one for the evolution of pure quantum states and another for the evolution of a density matrix describing a system coupled to reservoirs. Both methods are based on writing expansion coefficients of a state in terms of products of matrices, the so-called matrix product state (MPS) ansatz for pure states and matrix product operator (MPO) ansatz for density matrices. In the following, we will only briefly overview the tDMRG method. For a more complete presentation see original references.\cite{tDMRG}

For the determination of a profile evolution from a given initial state, a time-dependent Schr\"{o}dinger equation has to be solved,
\begin{equation}
  \label{eq:1}
  i \pd{}{t}\ket{\psi} = H \ket{\psi}.
\end{equation}
State vector $\ket{\psi}$, spanned by the tensor product of local Hilbert spaces for each site, can be written as
\begin{equation}
  \label{eq:2}
  \ket{\psi} = \sum_{\underline s} c_{\underline s} \ket{\underline s},
\end{equation}
where $\underline s$ enumerates all basis vectors of the Hilbert space $\underline s=(s_1,\dotsc,s_L)$ with $s_l \in \{0,1\}$ for the spin-down/up states. The dimension of Hilbert space for the spin chain of length $L$ is $2^L$. Solving the Schr\"{o}dinger equation using exact diagonalization is thus possible only for very small systems. However, it turns out that often not the whole Hilbert space is relevant for the solution and much larger systems can be solved by a clever choice of basis vectors, for instance, as is done in the tDMRG method. In MPS formulation, expansion coefficients $c_{\underline s}$ are expressed as traces of product of $L$ matrices $\bm A_l^{s_l}$, $l=1,\dotso,L$, of dimension $K \times K$,
\begin{equation}
  \label{eq:3}
  c_{\underline{s}}=\tr(\bm A_1^{s_1}\dotsm \bm A_L^{s_l}).
\end{equation}
Time evolution of the system is therefore described by time-dependent matrices $\bm A_l^{s_l}(t)$, which can be efficiently calculated if the propagator $U(t)=\exp{(-{\rm i} H t)}$ can be factorized into a product of unitaries that act only on two nearest-neighbor sites of a chain. This can be approximately done for short time-step propagation $U(\tau)$ using Suzuki-Trotter expansion. The system Hamiltonian is separated into two parts, $H=H_1 + H_2$, where all terms grouped inside each $H_1$ and $H_2$ mutually commute. General Suzuki-Trotter expansion can be written\cite{suzuki-trotter} as $U(\tau) = \prod_k \exp(-{\rm i} \alpha_k H_1 \tau) \exp(-i \beta_k H_2 \tau) + \mathcal{O}(\tau^{p+1})$, where the order $p$ depends on the actual scheme used. The $\mathcal{O}(\tau^{p+1})$ contribution gives rise to Suzuki-Trotter error, which can be reduced by either using expansions of higher order (having larger $p$) or reducing the step size $\tau$. After each time step, the MPS form of the state has to be restored using the SVD algorithm. In the process the matrices $\bm A_l^{s_l}$ are enlarged and to prevent an exponential growth of their dimension they must be truncated back to dimension $K$. This produces a truncation error $\epsilon_{\rm{trunc}}$ which depends on the entanglement of the state being described. A cumulative truncation error due to truncations at each time step can serve as a very rough estimate of the precision. A more robust method to check accuracy though is to simply increase $K$ and check that the results do not change. By using imaginary time step $\tau \rightarrow i \tau$, the tDMRG method can also be used for obtaining a ground state of a given Hamiltonian.

The tDMRG method for evolution of pure-states can be extended to the evolution of mixed states by introducing a $4^L$-dimensional Hilbert space of operators, with arbitrary operator given by
\begin{equation}
  \label{eq:4}
  \ket{\rho} = \sum_{\underline s} c_{\underline s} \ket{\sigma^{\underline s}},
\end{equation}
where $\sigma^{\underline s}=\sigma^{s_1}_1\dotsm\sigma^{s_L}_L$, $\underline s=(s_1,\dotsc,s_L)$, and $s_l \in \{0, 1, 2, 3 \}$, with $\sigma^0 = \mathds{1}$, $\sigma^1 = \sigma^x$, $\sigma^2=\sigma^y$, $\sigma^3=\sigma^z$, while lower indices denote a site along a chain. Following the reasoning of the MPS formulation for pure state evolution, coefficients $c_{\underline s}$ can be written in a MPO form, analogous to the MPS ansatz in Eq.~\eqref{eq:3}, with $s_l$ enumerating the basis of single-site density operators. The time evolution of density matrix $\rho$ is governed by the Lindblad equation, with a formal solution $\rho(t)=\exp(\mathcal{\hat L} t) \rho(0)$, where $\mathcal{\hat L}$ is a Liouvillian superoperator. Thus the propagator for time step $\tau$ can be again factorized as a product of local propagators using Suzuki-Trotter expansion, providing an efficient numerical scheme for the time evolution of a density operator. Because $\mathcal{\hat L}$ also contains dissipative terms due to a coupling with reservoirs, some care must be taken to ensure that one has a Schmidt-decomposed form of MPO at each step. Details of our implementation can be found in the appendix of Ref.~\onlinecite{NJP10}.

In the simulations of pure states, the Suzuki-Trotter expansion of second order was used ($p=2$), mostly with the time step of size $\tau=0.05$. The dimensions of the decomposition $K$ were chosen such that the truncation error on each time step during the evolution did not exceed $\epsilon_{\text{trunc}}=10^{-4}$ (required dimensions of decomposition $K$ were of the order of $\sim 100$). For master-equation simulations a fourth-order method with a time step $\tau=0.05$ has been used. The results were also verified by repeating simulations at somewhat higher decomposition sizes $K$ and smaller time steps, showing no deviation in results.

\section{Transport properties}
\label{sec:transport}
The main goal is to study magnetization transport in the Heisenberg model at low energies, particularly for $\Delta>1$, where magnetization transport looks diffusive at an infinite temperature.\cite{JSTAT09,Stein:09,PRL:11} We shall use two methods, one will be spreading of localized packets and the other the dependence of the spin current on the system size in the stationary nonequilibrium state of a chain coupled to different reservoirs at chain ends. It is useful to have a ``thermometer'' with which we will be able to determine the temperature to which respective simulations correspond. In the thermodynamic limit of stationary states of a master equation under weak driving the state is in a local quasiequilibrium and the temperature is a well defined concept. In the wavepacket simulations though, in which packets are initially far from equilibrium, we prefer to speak about energy density instead, as there is no local quasi-equilibrium and the temperature is not well defined. Another important scale in the gapped regime of $\Delta>1$ is the size of the energy gap between the ground and the first excited state.

In this section, we shall therefore first establish the values of the ground-state energy, the gap, and the relation between temperature and the energy density for the \textit{XXZ} Heisenberg model, in particular for $\Delta=1.5$ used later. Next, we shall present the results for the wave-packet simulations, followed by master-equation results.

\subsection{Ground state energy, the gap, and the temperature}
\begin{figure}[ht!]
\includegraphics[width=0.48\textwidth]{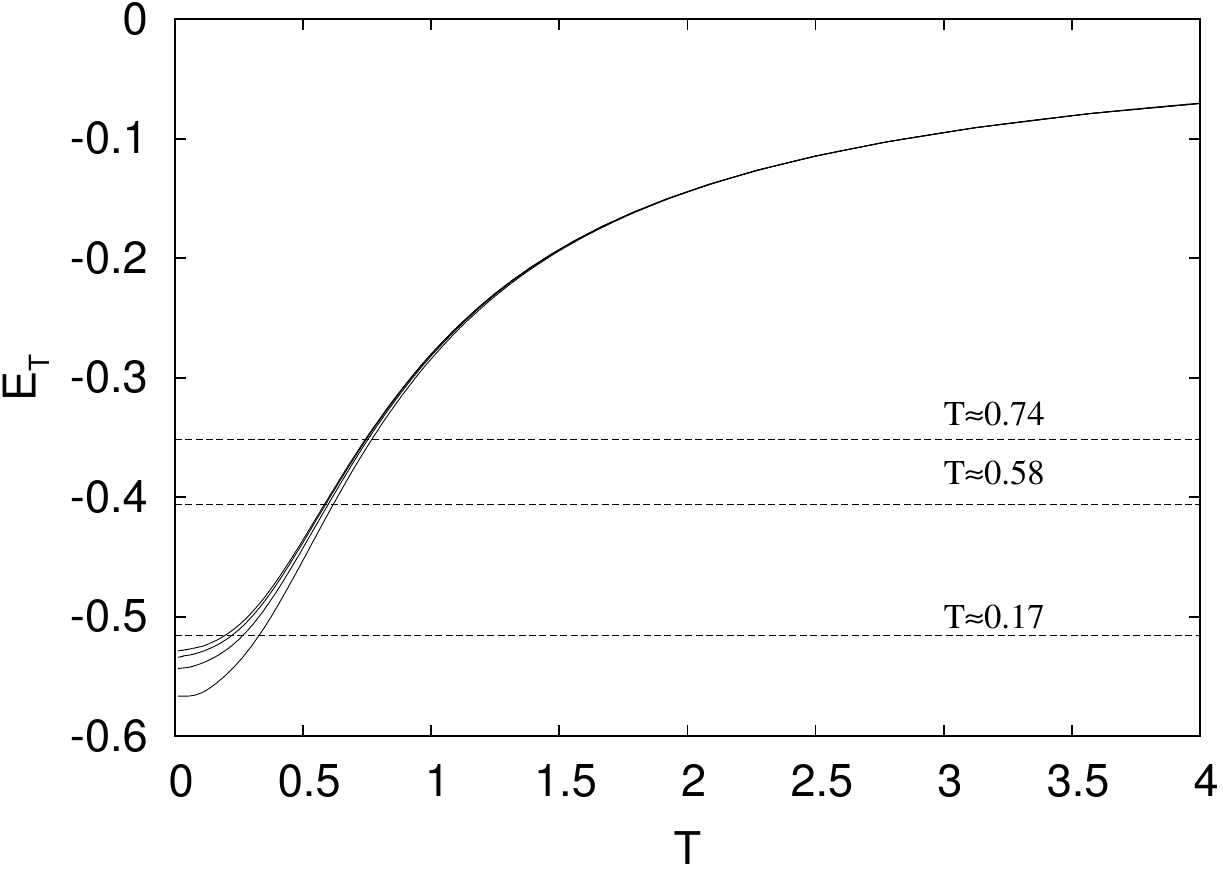}
\caption{Dependence of the canonical energy density $E_{T}$ (\ref{eq:ET}) on the temperature for the anisotropic Heisenberg model, $\Delta=1.5$, and open boundary conditions. Curves are for $L=8,16,32,64$, bottom to top, while horizontal lines denote the average energy densities of Gaussian packets we use to study transport (note that all are much higher than the gap).}
\label{fig:kanon}
\end{figure}
In order to be able to estimate where in the energy spectrum our initial states are, we have calculated the ground-state energy of the anisotropic Heisenberg model,\cite{Walker:59} the size of the gap, and for both quantities also finite-size corrections. Using numerically exact diagonalization for small sizes and an imaginary-time tDMRG for longer chains (up to $L=64$), we have determined by fitting that for anisotropy $\Delta=1.5$ and open boundary conditions the energy density $h_0$ in the ground-state (which is always from the sector with $M=0$) is
\begin{equation}
h_0=\frac{\ave{H}}{L-1}\approx -0.5234-\frac{0.29}{L-1}.
\label{eq:groundh}
\end{equation}
Here $\ave{H}$ is the expectation value in the ground state of $H$. Finite-size correction to the asymptotic energy density is therefore of the order ${\cal O}(1/L)$. At $L=200$, used in our simulations, the ground-state energy density is $h_0 \approx -0.5247$. Interesting to note is that for periodic boundary conditions finite size correction is smaller, namely, the ground-state energy density is $h_0^{\rm PBC}\approx -0.5234-0.89/L^2$. The gap between the ground state and the first excited state (within the same symmetry class) is for $\Delta=1.5$ and open boundary conditions
\begin{equation}
E_1-E_0\approx 0.068+\frac{7}{L}.
\label{eq:gap}
\end{equation}
For a chain of length $L=200$ the gap is $E_1-E_0 \approx 0.102$. As we shall see, the energies of our initial conditions will always be significantly above the ground state gap. At the energy scale below the gap, transport is trivially insulating and we are not interested in this regime. We have also calculated the relation between the thermodynamic temperature and the energy density $E_{T}$ in a canonical state,
\begin{equation}
E_{T}=\frac{\tr{(\rho_{T} H)}}{L-1},\qquad \rho_{T}=\frac{\exp{(-H/T)}}{\tr{\,\exp{(-H/T)}}}.
\label{eq:ET}
\end{equation}
Results are in Fig.~\ref{fig:kanon}. Such relation can be used as a ``thermometer''~\cite{PRE10}: for a given local energy density $E_l$, one can determine to what thermodynamic temperature this corresponds by equating $E_l=E_{T}$ and solving for the temperature $T$. One should be aware though that the validity of the canonical distribution is by no means granted for integrable systems. In fact, for specially chosen local reservoirs within the Lindblad master equation, deviations from the canonical distribution can be significant for integrable systems;\cite{PRE10} for discussion of open systems with general nonlocal coupling to reservoirs, see, e.g.,~Ref.~\onlinecite{Schaller10}.

\subsection{Localized packets}
\label{sec:transport_gaussian}

One way to study transport is to initiate a spin chain in an initial state that is nonequilibrium with respect to the Hamitonian generating time evolution only in a small localized region. In other words, one prepares a localized initial packet and then studies how such a packet spreads in time. To obtain a state that is out of equilibrium only in a small region of space, we put the chain into a spatially inhomogeneous external magnetic field $B_l$ and use the ground state of such a system as an initial state for our evolution.

\subsubsection{Preparation of initial states}
To prepare the initial state we take the Hamiltonian
\begin{equation}
  H_0=H+\sum_{l=1}^L S_l^z B_l,
  \label{eq:H_B}
\end{equation}
where $H$ is the \textit{XXZ} model Hamiltonian from Eq.~\eqref{eq:hamiltonian_XXZ}. For such Hamiltonian, the ground state $\ket{\psi_0}$ was obtained using imaginary-time tDMRG method. At time $t=0$ the magnetic field is then removed and the initial state evolves according to $H$. The actual spatial dependence of $B_l$ must be chosen in such way that time evolution of magnetization enables us to distinguish between ballistic and diffusive behavior. Its detailed form will be given further on a per-case basis.

Fermionic nature of the antiferromagnetic ground state of the \textit{XXZ} model leads to Friedel oscillations in magnetization profile $\braket{S^z_l}$ which get more pronounced when either effective interaction between fermions ($\Delta$) or magnetic disturbance for obtaining initial state gets larger. Friedel oscillations have a characteristic wavelength of $2\pi/(2k_{F})$, where $k_F=\pi/2$. As intensity of these oscillations can blur the effective magnetization profile, simple averaging over two neighboring sites was used in the majority of calculations, $\tilde S^z_l = (\braket{S^z_{l-1}} + \braket{S_{l}^z})/2$, similar to Ref.~\onlinecite{Langer2009}. The same averaging was also used for the energy density profiles $\tilde E_l = (\braket{E_{l-1}} + \braket{E_{l}})/2$, where the energy density operator is $E_l=\frac{1}{2}(S^+_lS^-_{l+1}+{\rm H.c.})+\Delta S_l^{z}S_{l+1}^{z}$.

Spatial dependence of the initial external magnetic field for the Gaussian-shaped initial profiles is given by
\begin{equation}
  \label{eq:Gaussian_magnetic_field}
  B_l=B e^{-\frac {[l-(L+1)/2]^2}{2 \sigma_B^2}} - B_{0},
\end{equation}
where $B$ determines the intensity of an initial disturbance, $\sigma_B$ its width and $B_0$ an overall offset of magnetic field that is used to adjust the total magnetization $M$ of spin chain.

\subsubsection{Evolution of magnetization profiles}
\begin{figure}[ht!]
\includegraphics[width=0.5\textwidth]{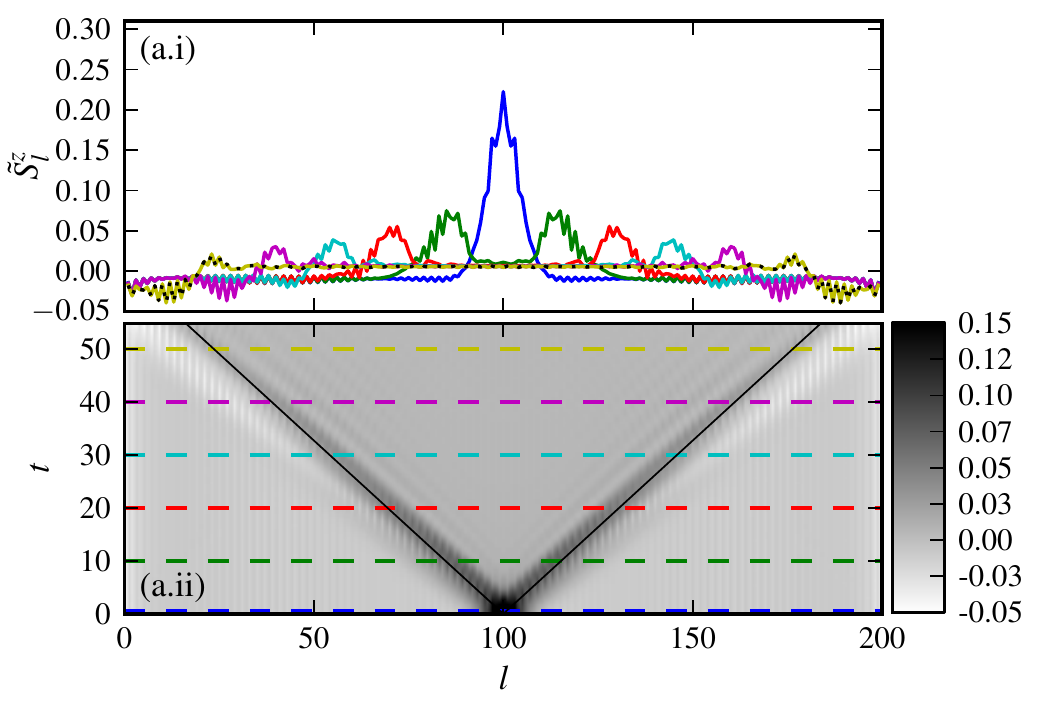}
\includegraphics[width=0.5\textwidth]{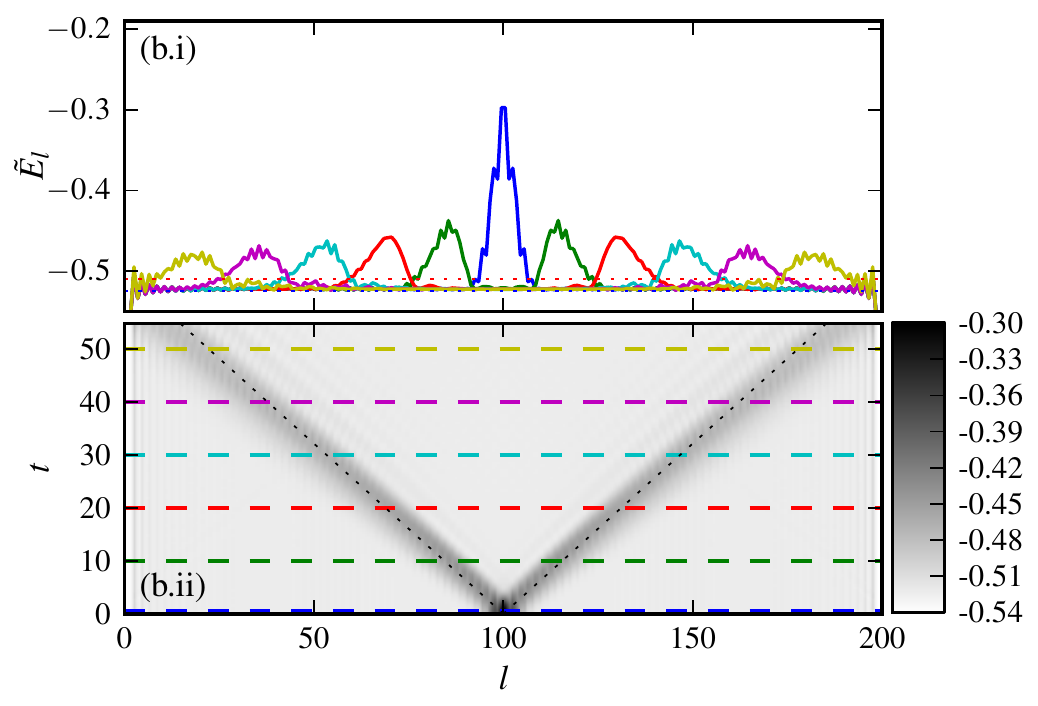}
\caption{(Color online) Time evolution of the initial Gaussian packet obtained with $B=2$, $\sigma_{B}=5$ and $B_0=-0.18$ ($M=0$) for a chain with $L=200$ spins and $\Delta=1.5$. Subfigures \mbox{(a.i)} and \mbox{(a.ii)} show the evolution of local magnetization, and subfigures \mbox{(b.i)} and \mbox{(b.ii)} the evolution of the energy density. In the gray-coded density plots \mbox{(a.ii)} and \mbox{(b.ii)} horizontal dashed lines denote times $t=0,10,20,30,40,50$ at which cross-sections are shown on the subfigures \mbox{(a.i)} and \mbox{(b.i)}. Two slanted lines in the spin density plot \mbox{(a.ii)} indicate ballistic spreading of the magnetization with the speed $v\approx 1.53$. Two dotted slanted lines in energy density plot \mbox{(b.ii)} indicate ballistic energy spreading with $v_{E} \approx 1.55$. Both $v$ and $v_E$ were determined by fitting the peaks of spin and energy profiles at various times. Two dotted horizontal lines in the plot of energy density profiles \mbox{(b.ii)} are the average energy density $h \approx -0.516$ of the initial state and of the ground state $h_0 \approx -0.525$. In subfigure \mbox{(a.i)} the dotted line, overlying $t=50$ cross-section, indicates spin profile calculated with MPS decomposition size $K=125$, demonstrating negligible deviations from the spin profile calculated at $K=90$ (underlying bold line) even at longest simulation times. }
\label{fig:sig5B2_S0}
\end{figure}
Instead of focusing on the growth of the packet's variance with time, as for instance in Ref.~\onlinecite{Langer2009}, we focus on the evolution of the whole magnetization profile.\cite{foot_sigma} The reason is that, in the variance one can get spurious effects as we shall discuss at the end of this subsection in~\ref{sec:B0}. In this section, we always take $\Delta=1.5$. We first pick a moderate $B=2$, packet width $\sigma_{B}=5$ and a compensating magnetic field $B_0=-0.18$, resulting in an initial state with $M=0$. In Fig.~\ref{fig:sig5B2_S0}, we show a density plot showing local magnetization along the chain for times up to $t=50$. At few time slices, we also show magnetization profiles. Two similar plots are also shown for the energy density. The energy $E=\langle \psi_0 | H | \psi_0 \rangle$ of the initial state $\ket{\psi_0}$ is about $1.8$ above the energy of the ground state, which is much more than the value of the gap that is $E_1-E_0 \approx 0.1$. The average energy density of our state is $h=E/(L-1)=-0.516$. In equilibrium, this would correspond to the canonical expectation at the temperature $T \approx 0.17$; see also Fig.~\ref{fig:kanon}. Note that we do not make any claim that the state is in or is close to being in local equilibrium, in which case one could use a local temperature. From the profile plots we can see that the two bumps, moving away symmetrically from the origin, spread ballistically. In the density plots these are visible as ballistic jets. The speed of the wave front is almost the same for the energy and magnetization spreading. Based on these results one would be tempted to conclude that the transport is ballistic. However, one should be aware that the statement about transport is an asymptotic one, that is, for long times. In relatively small chains available ($L=200$) it could well happen that we have not yet reached this asymptotic regime. In fact, looking at the magnetization profiles one can see that there is some nontrivial dynamics going on in the region between the bumps, and that the bumps change shape and height with time. Also, one can argue on general ground that the short-time dynamics for generic models is always ballistic. For an explicitly solvable quantum model, where a transition from a ballistic spreading at short times to a diffusive at large times can be shown, see Ref.~\onlinecite{Eisler:11}. We will in fact see that master-equation simulations, presented later, indeed support such a scenario. They also indicate that the transition time to asymptotic diffusive behavior might be very large for the small-energy wavepackets shown in Fig.~\ref{fig:sig5B2_S0}.

\begin{figure}[ht!]
  \includegraphics[width=0.5\textwidth]{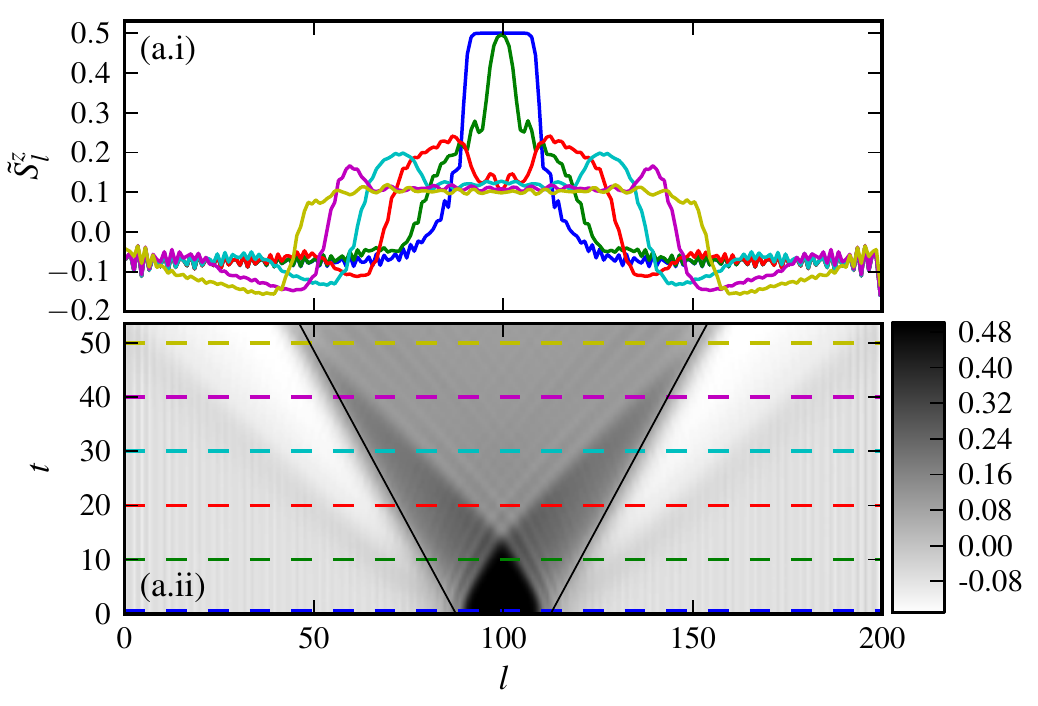}
  \includegraphics[width=0.5\textwidth]{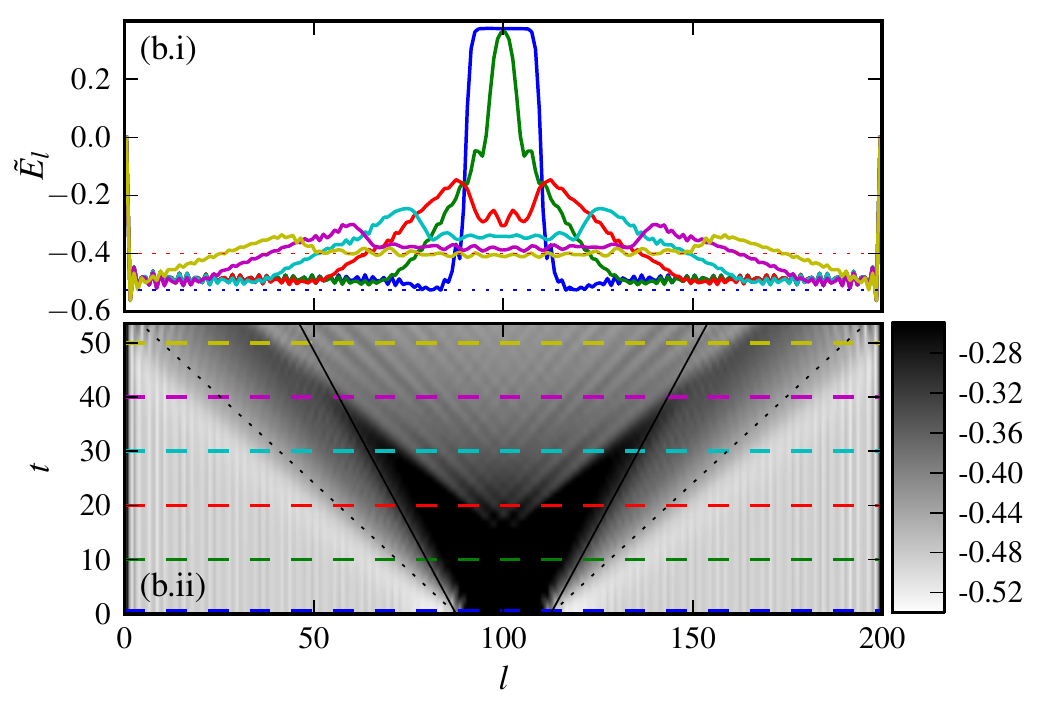}
  \caption{(Color online) Time evolution of the initial Gaussian packet obtained with  $B=5$ and $\sigma_{B}=10$, $L=200$, $\Delta=1.5$. Subfigures \mbox{(a.i)} and \mbox{(a.ii)} show the evolution of local magnetization, and subfigures \mbox{(b.i)} and \mbox{(b.ii)} the evolution of the energy density. In density plots \mbox{(a.ii)} and \mbox{(b.ii)} horizontal dashed lines denote times $t=0,10,20,30,40,50$ at which cross-sections are shown on the subfigures \mbox{(a.i)} and \mbox{(b.i)}. Two slanted lines in the subfigure \mbox{(a.ii)} indicate ballistic spreading of the magnetization with the speed $v\approx 0.77$, determined by fitting to the wavefront reaching $\tilde S_l=0.05$. Two dotted slanted lines in subfigure \mbox{(b.ii)} denote speed of energy spreading $v_E \approx 1.55$ as determined in Fig.~\ref{fig:sig5B2_S0}, while solid slanted lines indicate speed of spin profile wavefronts from subfigure \mbox{(a.ii)}. Two horizontal lines in the plot of energy density profiles are the average energy density of the initial state $h\approx -0.406$ and of the ground state $h_0\approx -0.525$.}
\label{fig:A10S10}
\end{figure}
\begin{figure}[ht!]
\includegraphics[width=0.5\textwidth]{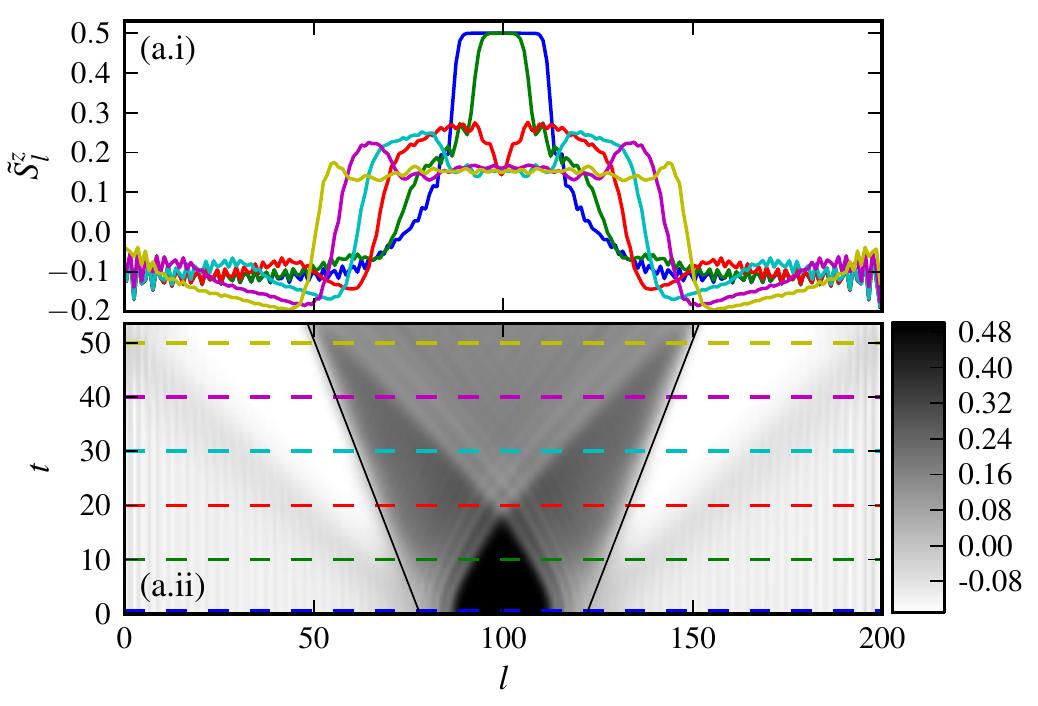}
\includegraphics[width=0.5\textwidth]{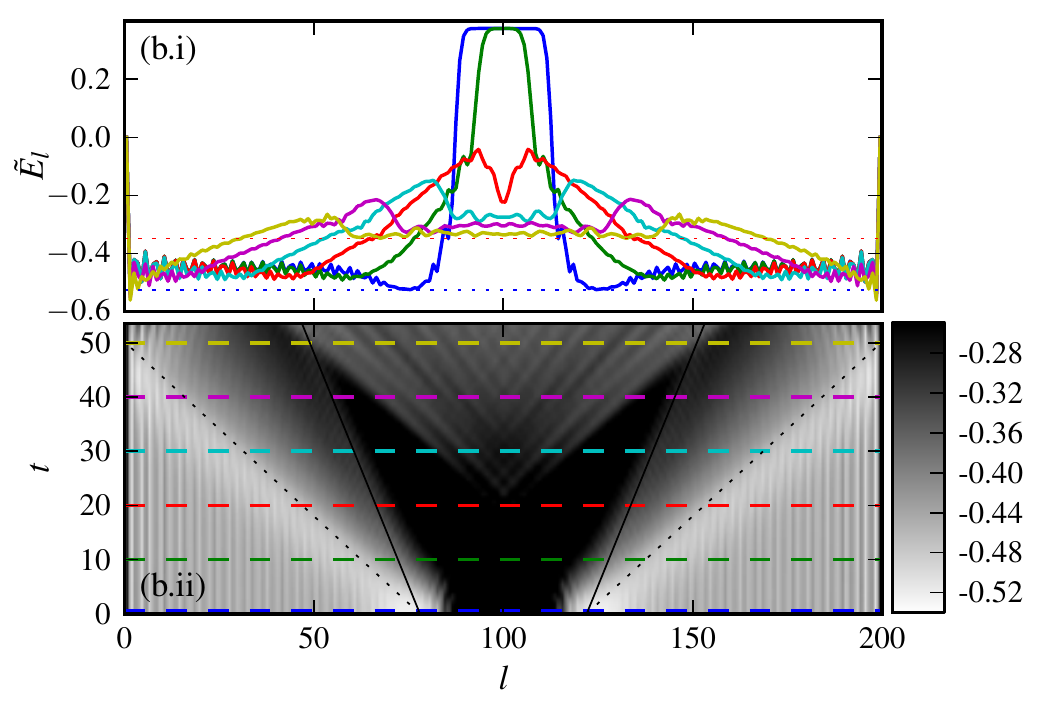}
\caption{(Color online) Time evolution of the initial Gaussian packet obtained with $B=5$ and $\sigma_{B}=15$, $L=200$, $\Delta=1.5$. Subfigures \mbox{(a.i)} and \mbox{(a.ii)} show the evolution of local magnetization, and subfigures \mbox{(b.i)} and \mbox{(b.ii)} the evolution of the energy density. In density plots \mbox{(a.ii)} and \mbox{(b.ii)} horizontal dashed lines denote times $t=0,10,20,30,40,50$ at which cross-sections are shown on the subfigures \mbox{(a.i)} and \mbox{(b.i)}. Two slanted lines in the subfigure \mbox{(a.ii)} indicate ballistic spreading of the magnetization with the speed $v\approx 0.55$, determined by fitting to the wavefront reaching $\tilde S_l=0.05$. Two dotted slanted lines in subfigure \mbox{(b.ii)} denote speed of energy spreading $v_E \approx 1.55$ as determined in Fig.~\ref{fig:sig5B2_S0}, while solid slanted lines indicate speed of spin profile wavefronts from subfigure \mbox{(a.ii)}. Two horizontal lines in the plot of energy density profiles are the average energy density of the initial state $h\approx -0.352$ and of the ground state $h_0\approx -0.525$.}
\label{fig:A10S15}
\end{figure}

To nevertheless be able to better assess the nature of transport also in packet-spreading simulations, we go to higher energies. There is strong numerical support that the spin transport at an infinite temperature is diffusive.\cite{JSTAT09,Stein:09,PRL:11} The ballistic spreading of jets at low energies, seen in Fig.~\ref{fig:sig5B2_S0}, should therefore weaken, for instance, slow down, or even entirely disappear, if the transition time to diffusive behavior gets short enough at higher energies. To verify this hypothesis we have simulated packets at higher energies by simply increasing the amplitude $B$ and the width $\sigma_{B}$ of the initial magnetic field. Note that in this way we also produce states that are more strongly nonequilibrium, at least for short times. In Fig.~\ref{fig:A10S10} we used $B=5$ and $\sigma=10$ ($B_0=-0.86$) to prepare the initial state in a sector with zero total magnetization. The energy density of such an initial state is $h=-0.406$ and is therefore significantly above the ground state ($\approx 240$ times the gap). This energy density would in equilibrium correspond to temperature $T \approx 0.58$. We can see from the results in Fig.~\ref{fig:A10S10} that the wavepacket front still spreads ballistically. There is one important difference though compared to the smaller-energy packet from Fig.~\ref{fig:sig5B2_S0}: the speed of the magnetization front is smaller, while the speed of the energy front is unchanged. Taking an even higher energy packet, obtained by $B=5$ and $\sigma_{B}=15$ ($B_0=-1.21$), the speed of the magnetization front decreases even more. Results for such a packet having an average energy density $h \approx -0.352$ (which would correspond to temperature $T \approx 0.74$) are in Fig.~\ref{fig:A10S15}.
From the spreading of packets at higher energies, several things can be concluded. First, in the regime of times and distances studied there is still a ballistic front in the magnetization that spreads at a constant speed. This speed decreases as one increases the energy, as it should in order to accommodate for a diffusive behavior at an infinite temperature. What happens with the front at larger times, for which one would need larger systems, is hard to infer. For the packet with the energy density $h=-0.516$ (Fig.~\ref{fig:sig5B2_S0}), the speed is $v \approx 1.53$, for the packet with the average energy density $h\approx -0.406$ (Fig.~\ref{fig:A10S10}), the speed is $v \approx 0.77$, and for the packet with $h \approx -0.352$ (Fig.~\ref{fig:A10S15}), the speed is $v\approx 0.55$. The second observation is that the speed of energy spreading is different than the speed of magnetization spreading. The speed of the energy front is always approximately $v_{E} \approx 1.55$, which is incidentally also equal to the speed of magnetization spreading at low energies\cite{foot_luttinger}. Note that the energy transport is ballistic in the Heisenberg model because the energy current is a conserved quantity. The third observation, especially visible in the spreading of the energy density in Fig.~\ref{fig:A10S15}, is that there is an energy dispersion. The shape of the energy packet changes with time. Parts with higher energy are ``overtaken'' by lower-energy excitations, visible as the stretching of the energy-density profile in front of the main peak.

All these simulations of the spreading of localized packets of magnetization show signs of ballistic spreading, either in terms of ballistic jets, or ballistically propagating wavefronts. One should be careful though about concluding that the transport is also ballistic at large times and in the thermodynamic limit, as these features could exhibit a nontrivial long-time behavior. As mentioned before, there could be a large time scale $\tau_{\rm diff}$, so that only for times larger than $\tau_{\rm diff}$ transport starts to show purely diffusive character. One can in fact connect this time scale with the diffusion constant $D$. Assuming for instance an exponential decay of the time-dependent spin current autocorrelation function, and taking into account that $D$ is equal to the integral of the autocorrelation function, one sees that the autocorrelation function decay time scales as $\propto D$. Therefore, if one has a diffusive behavior with a diffusion constant $D$, this introduces a characteristic diffusive time scaling as $\tau_{\rm diff} \sim D$. If diffusion constant $D$ is very large (we shall see that this is likely the case), one can have different behavior than the asymptotic diffusive one for $t \ll \tau_{\rm diff}$. Our simulations of wavepacket spreading therefore cannot distinguish truly ballistic transport from a diffusive with a large diffusive constant. To distinguish the two, one would have to look at the spreading of very wide and very shallow packets at long times, so that local deviations from equilibrium are small. Because with tDMRG one is limited to chains of few $100$ spins this limit can not be attained with the present computational resources. To nevertheless be able to say something about the spin transport in the anisotropic Heisenberg model at low energies, we also performed tDMRG simulations of transport in the master equation setting, where we can simulate systems at higher energies and some of the above mentioned problems do not appear.

Before doing that, we shall briefly present two interesting observations about the spreading of localized packets that are not directly related to transport. Readers interested only in transport properties can skip the next two subsections and jump directly to Sec.~\ref{sec:transport_master}.

\subsubsection{Short-time wavepacket pinch}
\begin{figure}[ht!]
  \includegraphics[width=0.5\textwidth]{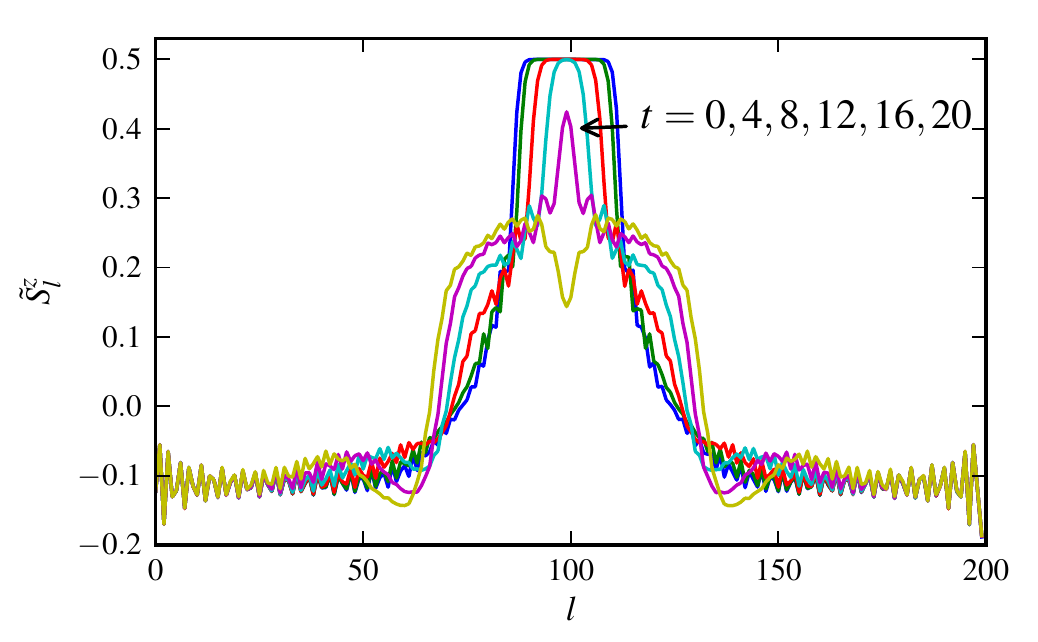}
  \caption{(Color online) Short-time pinch of the initial packet obtained with $B=5$ and $\sigma_{B}=15$ (same data as in Fig.~\ref{fig:A10S15}). After releasing the external magnetic field, the top of the packet first undergoes a contraction before it begins to spread at later times.}
  \label{fig:pinch}
\end{figure}

As an interesting side-remark, not directly related to our study of spin transport, at short times two wider packets in Figs.~\ref{fig:A10S10} and \ref{fig:A10S15} undergo a pinch. Their central part first contracts at very short times, and only then begins to spread. This can be seen in more detail in Fig.~\ref{fig:pinch} which shows short-time behavior of magnetization for the packet from Fig.~\ref{fig:A10S15} obtained with $B=5$ and $\sigma_{B}=15$. For times smaller than $\approx 20$ the central part of the packet shrinks while the shoulders expand.

\subsubsection{Magnetization offset $B_0$}
\label{sec:B0}
\begin{figure}[ht!]
  \includegraphics[width=0.5\textwidth]{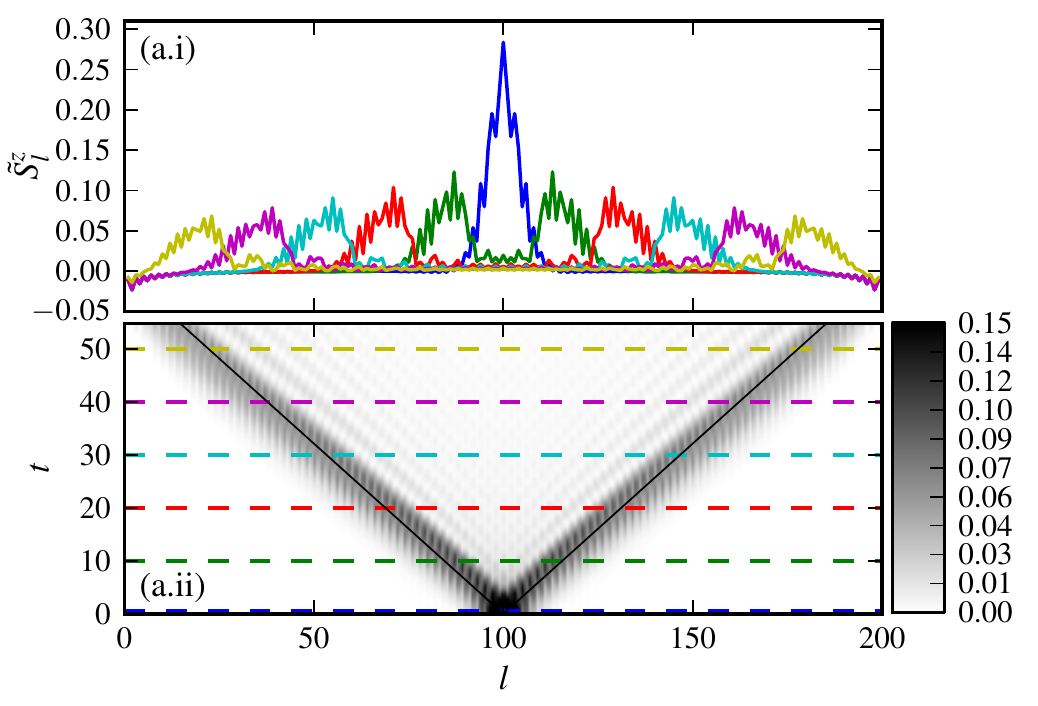}
  \includegraphics[width=0.5\textwidth]{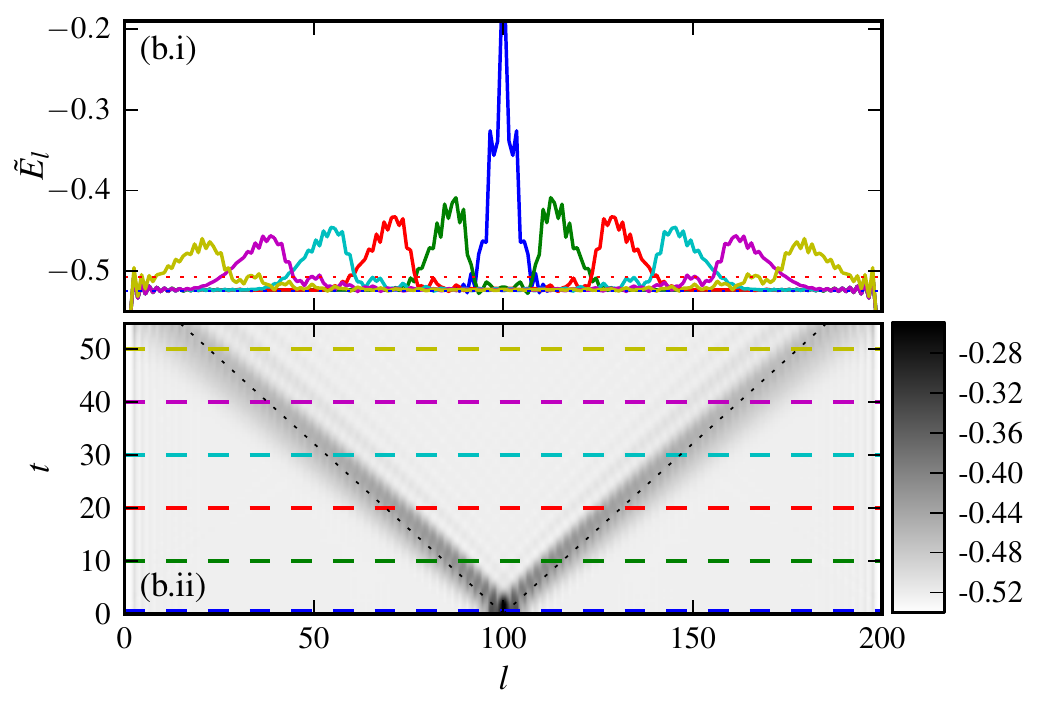}
  \caption{(Color online) Time evolution of the initial Gaussian packet obtained with $B=2$ and $\sigma_{B}=5$ without a compensating homogeneous magnetic field, $B_0=0$, thus having nonzero $z$-spin component $M = 2$. Subfigures \mbox{(a.i)} and \mbox{(a.ii)} show the evolution of local magnetization, and subfigures \mbox{(b.i)} and \mbox{(b.ii)} the evolution of the energy density. In density plots \mbox{(a.ii)} and \mbox{(b.ii)} horizontal dashed lines denote times $t=0,10,20,30,40,50$ at which cross sections are shown on the subfigures \mbox{(a.i)} and \mbox{(b.i)}. Two slanted lines in the spin-density plot \mbox{(a.ii)} indicate ballistic spreading of the magnetization with the speed $v\approx 1.55$. Two dotted slanted lines in energy-density plot \mbox{(b.ii)} indicate ballistic energy spreading with $v_{E} \approx 1.55$. Both $v$ and $v_E$ were determined by fitting the peaks of spin and energy profiles at various times. Two dotted horizontal lines in the plot of energy-density profiles \mbox{(b.i)} are the average energy density of the initial state $h\approx -0.507$ and of the ground state $h_0\approx -0.525$. }
\label{fig:sig5B2_S2}
\end{figure}

\begin{figure}[t]
  \includegraphics[width=0.5\textwidth]{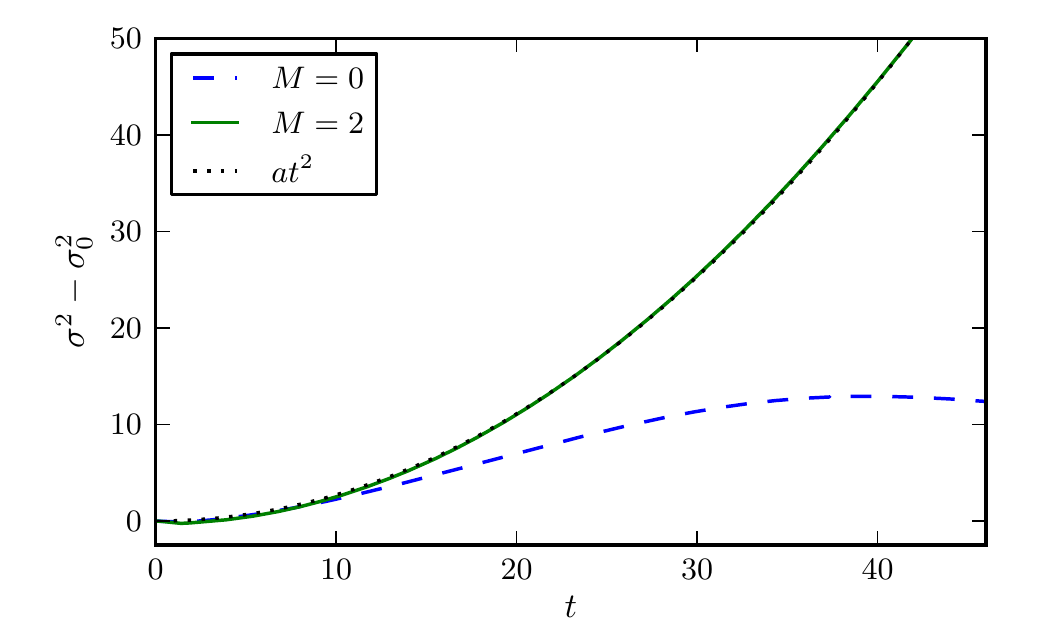}
  \caption{Time dependence of spin profile variance $\sigma^2(t) - \sigma^2_0$ for initial states obtained with $B=2$, $\sigma_B=2$ and total magnetization $M=0$, and $M=2$ (time evolution of corresponding profiles is shown on Figs.~\ref{fig:sig5B2_S0}~and~\ref{fig:sig5B2_S2}). Dotted line represents function $at^2$ fitted to the variance of $M=2$ case, demonstrating ballistic growth. The $M=0$ case on the other hand does not have simple time dependence of variance on attainable time scales. }
  \label{fig:variance}
\end{figure}

In this subsection we would like to make few comments about the influence of total magnetization of the spin chain $M$ on the evolution of spin profiles, specially on the variance $\sigma^2(t)$ of the profile. As $M$ is a conserved quantity of the \textit{XXZ} model, it is determined by the initial state, i.e., the values of the magnetic field $B$ and $B_0$ in Eq.~\eqref{eq:Gaussian_magnetic_field}. Remember that $B_0$ is used to tune the value of the total magnetization while retaining approximately Gaussian shape of the initial magnetization profile. In the linear response, transport properties of the \textit{XXZ} model are known to be related to the value of $M$. For states of nonzero magnetization, $M \neq 0$, ballistic behavior was predicted on the basis of Mazur's inequality as the spin current operator has a nonzero overlap with the conserved quantities of \textit{XXZ} Hamiltonian.\cite{Zotos1997} For $M=0$ this overlap is zero and no conclusion can be made, thus leaving the possibility of a diffusive transport. We generated initial states by two alternative choices of initial magnetic-field parameters $B$ and $B_0$. In the first case, total magnetization of the initial states was tuned to $M=0$ by using an appropriate compensating magnetic field $B_0$. In the second case, initial states were generated without a compensating magnetic field, $B_0=0$, thus resulting in a nonzero (but small) total magnetization $M \neq 0$. As system size approaches thermodynamical limit $L\rightarrow \infty$, there is no difference between the initial states generated by the above alternative choices of $B_0$ because $B_0 \rightarrow 0$. Yet, for the attainable system sizes $L \sim 200$ used in the numerical simulations, the details of the initial state preparation can have an observable effect on the dynamics. In the following we define the variance of the spin profile $\sigma^2(t)$ and summarize our observations about the influence of the initial state preparation on the variance and overall evolution.

Time dependence of variance, defined in the fermionic picture of the \textit{XXZ} model as
\begin{equation}
  \label{eq:variance_spin}
  \sigma^2=\frac{1}{n} \sum_{l=1}^L (l-\mu)^2 \cdotp
  \braket{n_l(t)}, \quad \mu=\frac{1}{n}\sum_{l=1}^L l \braket{n_l(t)},
\end{equation}
was extensively studied in Ref.~\onlinecite{Langer2009}. There, a ballistic spreading was established for $\Delta \leq 1$, while for $\Delta \gtrsim 1.5$ and $M=0$ a diffusive transport was advocated based on linear growth of $\sigma^2$ at intermediate times. For $M \neq 0$ it was shown~\cite{Langer2009} that one gets $\sigma^2 \sim t^2$. Linear growth of variance in the case of diffusive transport follows from the macroscopic diffusion law for magnetization transport, $\partial \tilde S^z (l)/\partial t = - D \partial^2 \tilde S^z (l)/\partial l^2$. We were able to confirm the ballistic growth of variance for $\Delta<1$. For $\Delta>1$, the situation is more complex. Depending on the details of the preparation of the initial state, more complex dynamics emerges, which can result in a linear variance growth for intermediate times~\cite{foot_sigma} (see Fig.~\ref{fig:variance} at $t \approx 10-25$). The actual shape of the central magnetization packet though does not follow the behavior that is expected from the macroscopic diffusion equation as can be seen in Fig.~\ref{fig:sig5B2_S0}.

While in the case of $\Delta<1$ the details of the initial state preparation do not have pronounced effects on the evolution (either by observing $\sigma^2(t)$ or the profile directly, both show a ballistic behavior), the effects are stronger in the case of $\Delta > 1$. For states with $B_0=0$ (Fig.~\ref{fig:sig5B2_S2}) the profile evolution as well as the time dependence of the variance (Fig.~\ref{fig:variance}) seem purely ballistic, at least at attainable time scales. For the $M=0$ states on the other hand, while the profile evolution still contains ballistic features (Fig.~\ref{fig:sig5B2_S0}), the variance does not grow as $\sigma^2 \propto t^2$ but instead has a nontrivial time dependence. While this nonballistic growth of variance may be taken as a hint of transition to a diffusive behavior, a detailed analysis of spin profiles reveals that observed behavior can be attributed to the negative magnetization ``bumps'' on the outer edges of spreading disturbance as seen on Fig.~\ref{fig:sig5B2_S0} for times $t>30$. As the value of the variance $\sigma^2(t)$ strongly depends on the profile values further away from the middle of the chain, such bumps have a strong impact on the variance. The occurrence of the bumps in the profile can be attributed to the emergence of a nonequilibrium magnetization plateau in the central region of the chain, between the packets traveling in opposite directions. Due to the conservation of total magnetization $M$, this nonequilibrium plateau of magnetization is compensated by an opposite deviation of magnetization on the edge of disturbance.
Variance of the profile therefore should not be taken as a sole criteria for the determination of transport regime. Thus we have focused mainly on the speed and the behavior of spin-profile wave fronts, emerging from the central peak, as these appear to be less dependent on the details of the initial state preparation. Note that we avoid making any definitive statement about the nature of transport from the wave-packet evolution only. These simulations will serve us only as a guide to master equation simulations that we present next.

\subsection{Master equation setting}
\label{sec:transport_master}

To induce a nonequilibrium steady state (NESS) we couple our spin chain to reservoirs. The coupling is described in an effective way via a set of Lindblad operators acting on the first two and the last two spins. The Lindblad master equation describing the evolution of the density matrix is~\cite{Lindblad,book}
\begin{equation}
\frac{{d}}{{d}t}{\rho}=\ii [ \rho,H ]+ {\cal L}^{\rm dis}(\rho)={\cal L}(\rho),
\label{eq:Lin}
\end{equation}
where the dissipative linear operator ${\cal L}^{\rm dis}$ is expressed in terms of Lindblad operators $L_k$,
\begin{equation}
{\cal L}^{\rm dis}(\rho)=\sum_k \left( [ L_k \rho,L_k^\dagger ]+[ L_k,\rho L_k^{\dagger} ] \right).
\end{equation}
To induce a NESS at a finite temperature we use Lindblad operators that couple to the first and last two spins of the chain -- the so-called two-spin bath. Details of the two-spin bath implementation can be found in Refs.~\onlinecite{JSTAT09,NJP10}. There are therefore 16 Lindblad operators $L_k$ at each end. They are chosen in such a way that they would induce a grandcanonical state $\sim \exp{(-H/T_{L,R}+\mu_{L,R} M)}$ on these two spins in the absence of Hamiltonian evolution by $H$. Because in our case the evolution by $H$ is present it introduces some boundary resistance effects that also affect the efficiency of the method. Due to these boundary effects it is rather difficult to cool the chain to very low temperatures~\cite{PRE10}. We use reservoirs with the same temperature $T_{L}$ at the left and $T_{R}$ on the right end of the chain, while the chemical potential is $\mu_{L}=0.1$ at the left and $\mu_{R}=-0.1$ at the right end. Because of this symmetric driving the average magnetization in the NESS is zero, as is also the energy current. To calculate $\rho(t)$, and therefore also NESS given by $\lim_{t \to \infty}\rho(t)$, we use the tDMRG method with a matrix product operator ansatz. After long time $\rho(t)$ converges to a stationary nonequilibrium state whose expectation values then give us transport properties. Due to boundary resistance the temperature in the bulk of the chain is not the same as the imposed temperature of the ``reservoir'' Lindblad operators. To determine the actual temperature in the system in the nonequilibrium steady state we use the expectation value of the energy density as a ``thermometer'',\cite{PRE10} equating it to the canonical one, and thereby determining the effective temperature, as described at the beginning of this section.

Because one has to simulate evolution of density operators instead of pure states, open system formulation is computationally more demanding than pure-state simulation. This typically means that somewhat smaller chains can be simulated. In addition, simulation at low energy (temperature) is more demanding because the operator-space entanglement of the NESS increases with decreasing temperature.\cite{Iztok:08} For instance, at an infinite temperature the NESS, being proportional to $\mathds{1}$, is separable with no entanglement. Therefore, due to computational constraints, we had to focus on somewhat higher energies than in the wavepacket simulations.

\begin{figure}[t]
\includegraphics[angle=-90,width=0.45\textwidth]{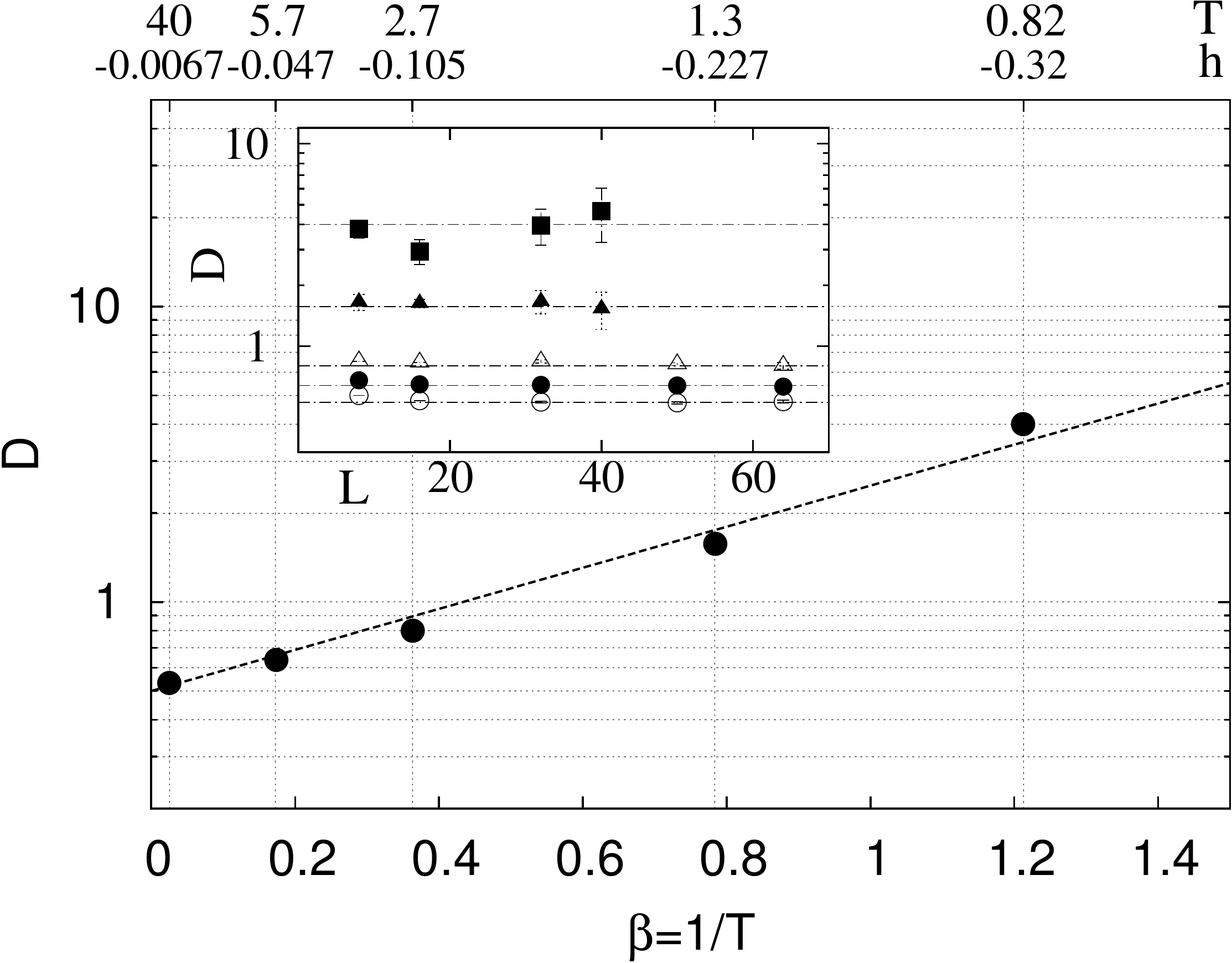}
\caption{Dependence of the diffusion constant obtained from NESSs on the inverse temperature. Dashed line suggests an exponential dependence on the inverse temperature, $D \approx 0.5 \exp{(1.6 \beta)}$. On the top axis we also list temperatures and energy densities corresponding to 5 NESSs. Inset: scaling of finite-size $D$ with the chain length $L$. Horizontal lines indicate the asymptotic $D$'s used in the main plot.}
\label{fig:kappa}
\end{figure}

Once we determine NESS for a given driving and length $L$ we calculate the expectation value of local magnetization, obtaining the difference in magnetization between left and right ends $\Delta S^z$, local spin current $j_l=(S_l^x S_{l+1}^y-S_l^y S_{l+1}^x)$ (which is independent of $l$), and local energy density. Because the deviation from the equilibrium zero magnetization is small, around $0.1$ in high energy simulations and $\approx 0.01$ at the lowest energy, and the imposed temperature is the same at both ends, the energy density in the NESS is constant along the chain. The effective temperature in the bulk of the chain is then determined by equating this energy density to the canonical one (local chemical potential in the NESS is close to zero). The finite-size diffusion constant can then be determined as $D=(L-1)\cdot j_l/\Delta S^z$. Taking into account definition of spin conductivity and of $D$ calculated here,\cite{pramana} the spin conductivity $\kappa$ can be obtained from the diffusion constant simply as $\kappa = \beta\, D$ ($\beta=1/T$). To ensure that the behavior is really diffusive and such $D$ does not depend on size $L$, we have calculated $D$ for sizes up to $L=64$. This data can be seen in the inset of Fig.~\ref{fig:kappa}. One can see that at high temperatures ($T > 2$, lower three chain lines in the inset) a good convergence of $D$ is obtained, signaling diffusive spin transport also at a finite temperature. At the lowest temperature $T \approx 0.82$ that we were able to simulate, convergence is less clear.
From the data on $D(T)$ in Fig.~\ref{fig:kappa}, one can clearly see that the diffusion constant increases at lower energies, the increase being perhaps exponential in the inverse temperature $\beta$, as indicated by a dashed line. At low temperatures/energies the diffusion constant can get very large. If we dare to extrapolate this dependence to the energy of the packet simulated in Fig.~\ref{fig:sig5B2_S0}, and we take the average temperature $T\approx 0.17$ as a crude estimate, the diffusion constant would be $D \sim 5000$. This means that to really observe a diffusive behavior in a wavepacket simulation one would have to simulate chain up to times of the order $\sim 5000$, demanding also chains of a similar size. We expect that ballistic jet, visible at short times, will disappear after this long time scale. All these results mean that with the tDMRG, being limited to few $100$ spins, one probably cannot conclusively say whether the spin transport is diffusive or ballistic at such low energies.

With an open-system tDMRG version we have nevertheless obtained strong indications that at not too low temperatures, $T > 1$, the anisotropic Heisenberg model is diffusive. The diffusion constant increases fast with decreasing temperature. Another thing we know is that at energy scales below the gap ($E_1-E_0\approx 0.1$ at $\Delta=1.5$), the system is trivially insulating; therefore, $D(T=0)=0$. At low temperatures that are still much above the gap there are then basically two possibilities: either the system is ballistic, meaning that the diffusion constant diverges at a finite temperature $T$, or the system is diffusive but with an exponentially large $D$. We find the latter scenario more plausible.

In any case, as one decreases $\Delta$ toward $1$, the gap disappears and therefore also an insulating state at $T=0$. At $\Delta=1$ the diffusion constant is therefore infinite at zero temperature (in agreement with a nonzero Drude weight~\cite{Shastry:90}), the transport therefore being either ballistic or anomalous. Recent results in Ref.~\onlinecite{PRL:11} show that at an infinite temperature and $\Delta=1$ it is anomalous (superdiffusive).

\section{Domain wall dynamics}
\label{sec:domain_walls}
In the present section we will focus on particular initial states whose time evolution is quite different from the one of the Gaussian packets. These are states with a domain-wall-shaped initial magnetization profiles. Such states are nongeneric and therefore do not influence our conclusions about magnetization transport reached earlier. The purpose is just to point out that, due to symmetry, there are particular states with different behavior. Such states were investigated in the context of domain-wall dynamics,\cite{Gobert2005, Antal1999} relaxation dynamics,\cite{Mossel2009} and domain-wall stability.\cite{Gochev1977,Gochev1983, Yuan2007} For a gapless regime ($\Delta<1$), numerical and analytical results suggest ballistic spreading of the initial domain-wall. Less is known about dynamics in a massive phase ($\Delta>1$). Time evolution of completely polarized domain walls have been studied numerically in Ref.~\onlinecite{Gobert2005} and analytically using semiclassical approximation in Ref.~\onlinecite{Lancaster2010}. Here we extend analysis to low-energy partially polarized domain states.

We get domain-wall initial states as ground states of the Heiseneberg \textit{XXZ} chain defined by Eqs.~\eqref{eq:hamiltonian_XXZ} and \eqref{eq:H_B} in a step-shaped magnetic field given by
\begin{equation}
  \label{eq:step_function}
  B_l = 2 B \Theta(\frac{L+1}{2}-l) - B,
\end{equation}
where $\Theta$ is Heaviside step function and $B$ is the magnitude of an external magnetic field that polarizes spins on the left and right side of the chain in opposite directions. In the limit $B \rightarrow \infty$, ground state of the chain is given by the simple product state $\ket{\uparrow...\uparrow \downarrow...\downarrow}$, while it is more complicated for finite $B$. Nonetheless, the shape of the spin profile $\braket{S^z_l}$ retains approximately step-like shape if we employ the nearest-neighbor averaging of $\tilde S_l^z$ to account for Friedel oscillations. The main question we studied was whether the initial domain-wall stays localized or it decays with time. To access this we have determined the domain-wall width $w$ at time $t$ as the distance between locations, positioned symmetrically around the middle of the chain, where the value of magnetization $\tilde S_l^z$ exceeds positive/negative offset value of magnetization $\pm M_w$ equal to half of the maximal one.

\begin{figure}[t!]
  \includegraphics[width=0.5\textwidth]{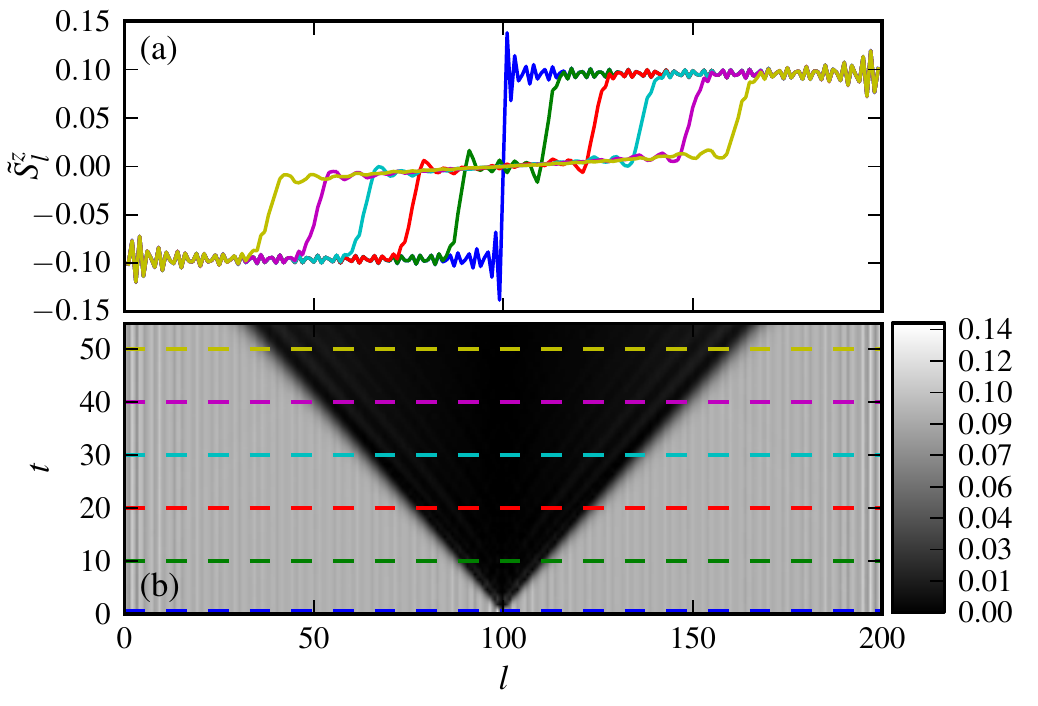}
  \caption{(Color online) Evolution of spin profile of a partially polarized domain-wall-like initial state in a gapless regime, $\Delta=0.5$ and $B=0.5$. In density plot \mbox{(b)}, horizontal dashed lines denote times $t=0,10,20,30,40,50$ at which cross-sections are shown on the subfigure \mbox{(a)}. Density plot (b) shows absolute value of spin profile $|\tilde S_l^z|$. Ballistic spreading of the domain wall is clearly seen.}
  \label{fig:stepD0.5B0.5}
\end{figure}

For the gapless case $\Delta<1$, the ballistic spreading of the domain wall was obtained by directly observing linearly increasing width of the domain wall ($w \sim t$). Ballistic behavior of domain-wall states was observed independently of the level of polarization, determined by initial state polarization $B$. An example of profiles for $B=0.5$, $\Delta=0.5$ is shown in Fig.~\ref{fig:stepD0.5B0.5}.

\begin{figure}[t!]
  \includegraphics[width=0.5\textwidth]{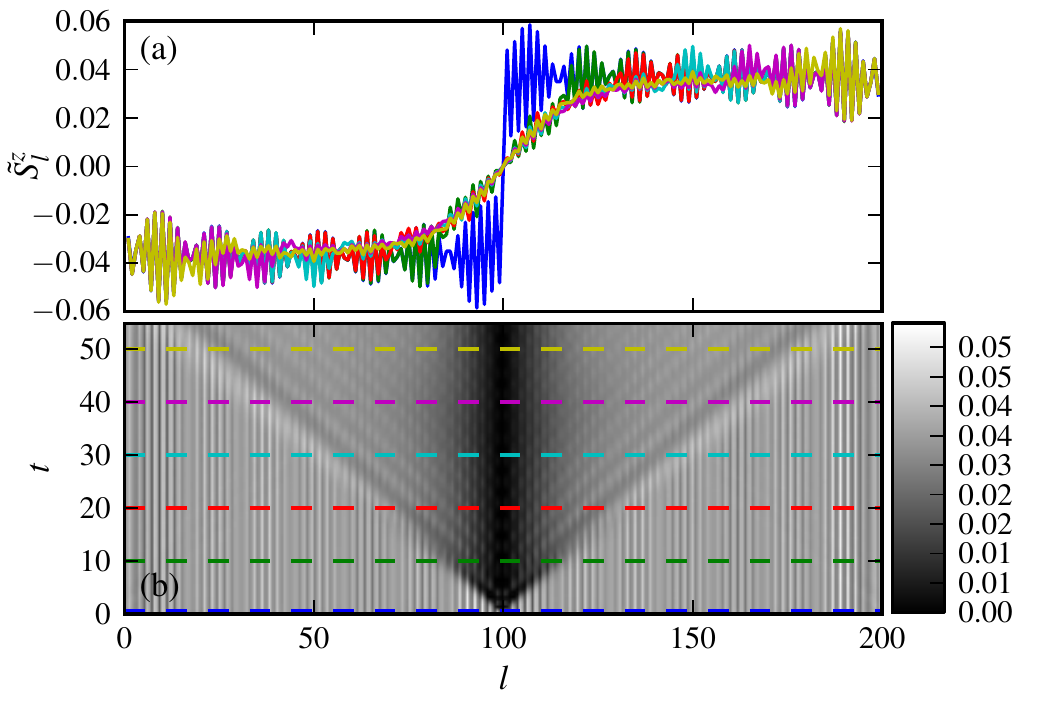}
  \caption{(Color online) Evolution of spin profile of a partially polarized domain-wall-like initial state in a gapped regime, $\Delta=1.5$ and $B=0.5$. In density plot \mbox{(b)} horizontal dashed lines denote times $t=0,10,20,30,40,50$ at which cross sections are shown on the subfigure \mbox{(a)}. Density plot (b) shows absolute value of spin profile $|\tilde S_l^z|$. Here the spreading of the domain wall stops and the domain wall starts to oscillate around a stable shape (see, e.g., cross sections at times $t=40$ and $t=50$).}
  \label{fig:stepD1.5B0.5}
\end{figure}

\begin{figure}[ht!]
  \includegraphics[width=0.5\textwidth]{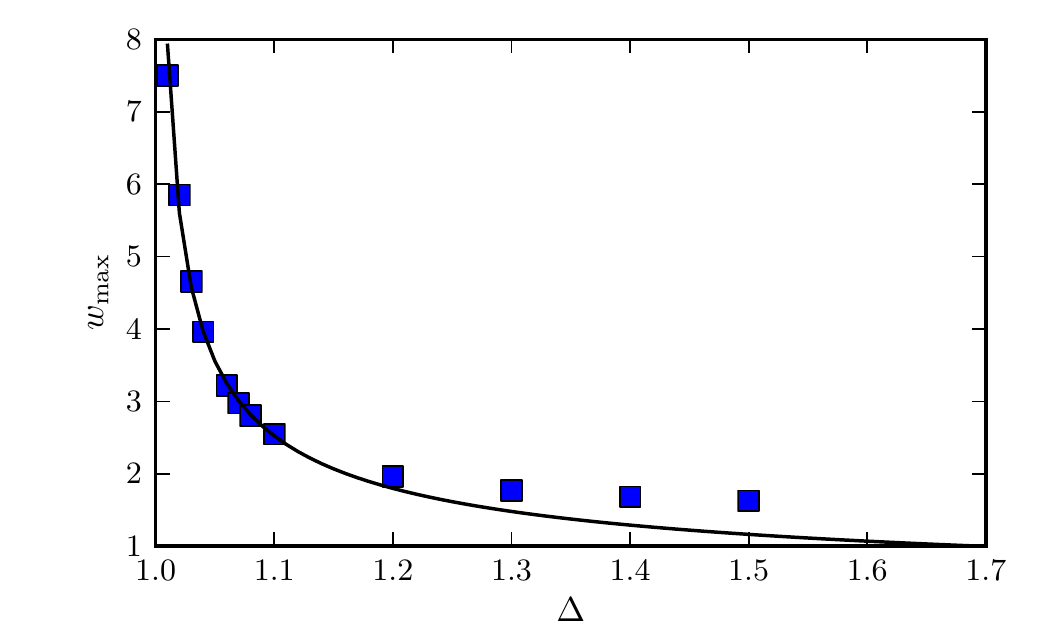}
  \caption{(Color online) Dependence of the maximal domain-wall width $w_{\rm{max}}$ on the anisotropy parameter $\Delta$ for $\Delta>1$ and the initial state of a completely polarized domain wall. Solid line has the functional dependence of Eq.~(\ref{eq:gochev}) with the best fitting $A=1.12$.}
  \label{fig:domain_widths}
\end{figure}
In the gapped regime $\Delta>1$, the dynamics of the initial domain wall is more complex. To determine whether the domain wall spreads diffusively or remains frozen due to localization, we have observed time evolution of the domain-wall width $w(t)$ for different values of $\Delta>1$ and for various polarizations of initial states, determined by $B$. For all $\Delta$, the initial state used has been the same and was obtained as a ground state of Hamiltonian with $\Delta=1$. At $t=0$ magnetic field $B$ was turned off and $\Delta$ was quenched to the target value. We have observed that for all $\Delta>1$, independently of the initial state polarization $B$, the width of the domain wall $w(t)$ grows until it attains some maximal value $w_{\text{max}}$ and then starts oscillating around some final value $w_{\text{final}}$ (see Fig.~\ref{fig:stepD1.5B0.5} for an example of $B=0.5$). Whether these oscillations are damped out as the time progresses could not be determined on the accessible time scales. It therefore appears that for $\Delta>1$ the initial domain wall stays localized. The functional dependence of the domain-wall width on the value of anisotropy $\Delta$ was suggested in Refs.~\onlinecite{Gochev1977,Gochev1983} for the ferromagnetic 1D Heisenberg model, having the form
\begin{equation}
  \label{eq:gochev}
  w=\frac{A}{\log(\Delta +
      \sqrt{\Delta^2-1})},
\end{equation}
where $A$ is dependent on the used definition of the domain-wall width. See also related interface
ground states in the ferromagnetic Heisenberg system with kink boundary terms in Ref. \onlinecite{kink}. Recently, localization of a fully polarized domain wall in the quantum anisotropic Heisenberg spin chain (ferromagnetic and antiferromagnetic) has been observed~\cite{ndc-epl} in the context of negative differential conductivity and explained~\cite{ndc} in terms of a one-magnon localization. See also recent work in Ref.~\onlinecite{Haque}. We have compared the measured maximal widths\cite{foot1} $w_{\text{max}}$ to the suggested scaling form, Eq.~(\ref{eq:gochev}). Results reported in Fig.~\ref{fig:domain_widths} are found to be in good agreement for large initial domain-wall polarizations ($B \rightarrow \infty$), especially around $\Delta \approx 1$, where domain-wall widths span multiple spin sites. At larger anisotropies the domain wall widths are in a range of a few spins so $w_{\text{max}}$ is strongly dependent on the details of the procedure for the domain width determination (e.g. averaging of magnetization profile and choice of domain-wall boundaries $M_w$), so that the measured value of the domain width deviates from the suggested scaling form. For weaker polarizations of initial states ($B \sim 1$) the determination of the domain-wall width $w_{\text{max}}$ is more involved as transient effects of the domain-wall dynamics get more pronounced. The functional dependence of $w_{\text{max}}$ for $B\sim 1$ was therefore not compared to the scaling form of Eq.~\eqref{eq:gochev}. However, we have noticed that the domain wall stays localized even for weakly polarized initial states, at least up to the timescales of $t\sim 100$.

\section{Conclusion}
We have studied magnetization transport at finite temperatures in the one-dimensional anisotropic Heisenberg model. By a combination of wavepacket spreading and the study of nonequilibrium steady states we reach a conclusion that the transport is diffusive in the gapped regime. By lowering the temperature the diffusion constant increases, perhaps exponentially so with the inverse temperature. A very large diffusion constant at low temperatures introduces a very long time and space scale that governs the transition from a ballistic behavior at short time to an asymptotic diffusive at long times. Using existing numerical techniques, that are limited to short times and small system sizes, it is therefore very difficult to observe diffusive behavior at very low energies.

\begin{acknowledgements}
We would like to thank T.~Prosen and P.~Prelov\v sek for discussions. Support by the Program P1-0044 and the Grant J1-2208 of the Slovenian Research Agency, the project 57334 by CONACyT, Mexico, and IN114310 by UNAM is acknowledged.
\end{acknowledgements}


\begin{thebibliography}{10}
\bibitem{foot_luttinger} The speed of the energy wavefront $v_E\approx 1.5$, as well as the speed of the spin wavefront at low energy, is close to the Luttinger liquid speed $u$ of low energy excitations at the isotropic point of the Heisenberg model, $u=\pi/2$. For details on Luttinger liquid treatment of the Heisenberg XXZ model see Ref.~\onlinecite{Giamarchi}.
\bibitem{foot1} {Maximal width of the domain wall during time evolution of the initial state is compared to the scaling limit because the final width $w_{\text{final}}$ requires much longer simulation times, which are not accessible for the cases of $\Delta \approx 1$. The use of maximal domain wall widths can be justified by observation that (in the cases studied) the functional dependence of maximal widths on anisotropy $\Delta$ is similar to functional dependence of final widths $w_{\text{max}}\sim w_{\text{final}}$}.
\end{thebibliography}
\end{document}